%% file: Manuscript.tex
\documentclass[twocolumn,hidelinks,10pt]{article} %amsart
\usepackage{graphicx,stfloats}
\usepackage{color}

 \usepackage{latexsym}
 \usepackage{subfigure}
 \usepackage{float}
 \usepackage{multirow}
 \usepackage{threeparttable}
 \usepackage[colorinlistoftodos,prependcaption,textsize=tiny]{todonotes}
 \usepackage{soul}
 \usepackage[version=4]{mhchem} 
 \usepackage{hyperref}
 \usepackage{placeins}
\usepackage{geometry}
\usepackage{amssymb}
\usepackage{siunitx}   % Einheiten
\usepackage{bm} 	   % bold Mathfonts
\usepackage[square,comma,sort&compress, numbers]{natbib}

\newcommand{\blue}{\color{black}}
\newcommand{\blueF}{\color{black}}

\bmdefine\u{u}
\bmdefine\btau{\tau}

\newcommand{\Institutes}{
\STFS Institute for Simulation of Reactive Thermo-Fluid Systems (STFS),\\Technische Universität Darmstadt, Otto-Berndt-Str. 2, Darmstadt 64287, Germany\\
\DLR Institute of Combustion Technology, German Aerospace Center (DLR),\\ Pfaffenwaldring 38-40, 70569 Stuttgart, Germany\\
\KIT Engler-Bunte-Institute, Division of Combustion Technology,\\Karlsruhe Institute of Technology Kaiserstr. 12, 76131 Karlsruhe, Germany}
\newcommand{\STFS}{\textsuperscript{1}}
\newcommand{\DLR}{\textsuperscript{2}}
\newcommand{\KIT}{\textsuperscript{3}}
\date{}
\usepackage[symbol]{footmisc}

\title{Numerical Investigation of the Local Thermo-Chemical State in a Thermo-Acoustically Unstable Dual Swirl Gas Turbine Model Combustor}

\author{ \textbf{T. Jeremy P. Karpowski\STFS ,   Federica Ferraro\textsuperscript{1}\footnote{shared first and corresponding author (ferraro@stfs.tu-darmstadt.de)}, Matthias Steinhausen\STFS, }\\ \textbf{Sebastian Popp\STFS, Christoph M. Arndt\DLR, Christian Kraus\KIT, }\\ \textbf{Henning Bockhorn\KIT, Wolfgang Meier\DLR, Christian Hasse\STFS}\\\\
\Institutes\\\\
ASME Turbo Expo 2022 Turbomachinery Technical Conference and Exposition\\
June 13-17 Rotterdam, \url{https://doi.org/10.1115/GT2022-83810}
}

\begin{document}
\maketitle

{\it
\noindent  In this work, the thermo-acoustic instabilities of a gas turbine model combustor, the so-called SFB606 combustor, are numerically investigated using Large Eddy Simulation (LES) combined with tabulated chemistry and Artificial Thickened Flame (ATF) approach. \blueF 
The main focus is a detailed analysis of the thermo-acoustic cycle and the accompanied equivalence ratio oscillations and their associated convective time delay. In particular, the variations of the thermo-chemical state and flame characteristics over the thermo-acoustic cycle are investigated. \color{black}
%The main focus is a detailed analysis of the thermo-acoustic cycle and its feedback mechanism due to equivalence ratio oscillations coupled with convective time delay. In particular, the variations of the thermo-chemical state and flame characteristics over the thermo-acoustic cycle, are investigated. 
\blueF
\noindent For the operating point flame B ($P_{th}=25kW$, $\Phi=0.7$), the burner exhibits thermo-acoustic instabilities with a dominant frequency of \SI{392}{\hertz},  the acoustic eigenmode of the inner air inlet duct. These oscillations are accompanied by an equivalence ratio oscillation, which exhibits a convective time delay between the injection in the inner swirler and the flame zone.
\color{black}
%\noindent This work investigates the variation of the thermo-chemical state over the thermo-acoustic cycle in flame B.
%\noindent A prior analysis is conducted for an adiabatic and an enthalpy-dependent reduced chemistry manifold to assess the influence of heat losses on the mean and phase-conditioned temperature measured in the experiments. The enthalpy-dependent reduced chemistry manifold is found to better predict the experimental data, indicating a non-negligible effect of the heat losses for the flame B conditions.  
% It is shown, that an adiabatic table results in deviations from the true thermo-chemical state which vary over the thermo-acoustic cycle. In contrast, the enthalpy-dependent table is able to capture the experimentally observed thermo-chemical state over the entire acoustic cycle.
\blue
\noindent   Two LES, one adiabatic and one accounting for heat losses at the walls 
by prescribing the wall temperatures from  experimental data and Conjugated Heat Transfer (CHT) simulations,
are conducted.
Results with the enthalpy-dependent table are found to predict the time-averaged flow field in terms of velocity, major species, and temperature with higher accuracy than in the adiabatic case. \blueF Further, they indicate, that heat losses should be \mbox{accounted} for to correctly predict the flame position. \color{black}
Subsequently, the thermo-chemical state variations over the thermo-acoustic cycle for the enthalpy-dependant case are analyzed in detail and compared with experimental data in terms of phase-conditioned averaged profiles and conditional averages. An overall good prediction is observed, although an overestimation of the oscillation amplitude yields a slight over-prediction of the velocity field in the low-pressure phases.
The results provide a detailed quantitative analysis of the thermo-acoustic feedback mechanism of this burner.
%Further, they indicate that the LES with a non-adiabatic tabulated chemistry model is required to predict the unsteady flow field and flame dynamics with an adequate level of accuracy. 
%Nonetheless, the enthalpy-dependent LES was found to be viable tool for the analysis of thermo-acoustics.\color{black}
}
\vspace{-10pt}
\section{Introduction}
\input{introduction.tex}

\section{Experimental Setup}
\input{experimental.tex}

\section{Numerical Model}
\input{numerics.tex}
\section{Numerical Setup}
\input{numerical_setup.tex}
\vspace{-3mm}
\section{Results}
\input{results.tex}
\vspace{-3mm}
\section{Conclusions}
\input{conclusion.tex}

\vspace{-5mm}
\section{Acknowledgements}

\noindent Calculations for this research were conducted on the Lichtenberg high-performance computer at TU Darmstadt.  This research was part of the "Center of Excellence in Combustion", which received funding from the European Union’s Horizon 2020 research and innovation program under grant agreement No$^\circ$ 952181. 
%\vspace{-3mm}
%\clearpage

\bibliographystyle{unsrtnat_mod}
\bibliography{Litrature}

%\printbibliography

\clearpage

\appendix
\end{document}

%% file: introduction.tex
To achieve high efficiency and low emission levels, modern gas turbines (GTs) are designed with swirl burners operating in the lean premixed or partially-premixed combustion regime~\cite{Huang2009, Gicquel2012}.
These burner configurations are, however, prone to hydrodynamic instabilities, such as Precessing Vortex Cores (PVC)~\cite{Oberleithner2015, Huang2009}, and thermo-acoustic instabilities, that may affect the combustor operations and even damage GT components~\cite{Poinsot2017}. 
%Several  studies investigated the mechanisms yielding to thermo-acoustic feedback loops. 
\blue
Multiple thermo-acoustic feedback loop mechanisms have been identified and discussed in several studies. % (see e.g.~\cite{Lieuwen2006}).   
%Main mechanisms are: i)
Acoustic modes can induce oscillations on the flame front kinematics, which can lead to variations of the flame surface~\cite{Schuller2003} and the flame \mbox{positions}~\cite{Lieuwen2006} resulting in heat release rate oscillations.
\mbox{Thermo-acoustic} modes can also be excited by vortex shedding as discussed in~\cite{Schadow1992,Palies2011}. Hydrodynamic instabilities and PVC may affect the thermo-acoustic cycle as well as the flame shape, as shown in~\cite{Boxx2012, Stoehr2018}.  Furthermore, pressure oscillations in the air or fuel supply can yield equivalence ratio oscillations: they are convected into the flame zone causing heat release rate fluctuations~\cite{Lieuwen1998,Meier2007}. 

%A detailed description can be found in~\cite{Lieuwen2006}.
%hydrodynamic instabilities and  the PVC, analysed in e.g.,~\cite{Boxx2012, Stoehr2018}, can affect the thermo-acoustic cycle as well as the flame shape.     

%kinematic effects such as flame movement or wrinkling induced by velocity fluctuations~\cite{Cosic2015} or equivalence ration fluctuations\cite{Meier2007}, which furthermore can influence each other~\cite{Kim2010}. 
\color{black}
Although significant progress has been made in the understanding and prediction of combustion instabilities~\cite{Gicquel2012}, accurately estimating their amplitude and frequency remains a challenge~\cite{Poinsot2017}.
%, due to the strong interaction of the acoustic modes with the turbulent flow and mixing field, turbulence-chemistry interaction, and the heat release rate.
\blue
To enhance the understanding of combustion instabilities, gas \mbox{turbine} model combustors (GTMCs)  have been intensively studied.  
Kim et al.~\cite{Kim2010} experimentally examined the response of a swirl-stabilized partially-premixed flame subjected to acoustic velocity and equivalence ratio oscillations. They found a dependency in the phase difference between the two perturbations determining the  linear or non-linear flame response. Non-linear response in swirl stabilized flames are studied in \cite{Cosic2015} for various oscillation amplitudes and a flame transfer function model was developed to determine the limit cycle behavior. It was observed that a saturation of the equivalence ratio perturbation is achieved already at relatively small amplitudes due to the increased turbulent mixing. 
The DLR dual swirl burner~\cite{Weigand2006a, Meier2006} and the PRECCINSTA burner \cite{Weigand2006, Meier2007} 
%and the SFB606 Gas Turbine Model Combustor~\cite{Arndt2015, Meier2016, Kraus2016, Arndt2017,Kraus2017} 
are further examples of partially-premixed swirl GTMCs designed to investigate combustion instabilities, which have been used as benchmark configurations for numerical simulations and model development in numerous studies. % due to the availability of  high-fidelity laser measurements for velocity, temperature and species mass fractions.  

 %Phase-averaged quantities 
%Although the thermo-acoustic feedback mechanism due to equivalence ratio oscillation and convective time delay has been qualitatively reproduced in~\cite{Franzelli2012, Lourier2017}, there has still been no detailed quantitative analysis of the flow field and flame characteristics over the oscillation phases. 
 
In the last decades, Large Eddy Simulations~(LES)  have become a fundamental resource for understanding complex three-dimensional turbulent reacting flows in combustors~\cite{Raman2019}, both in stable and unstable regimes, see e.g.~\cite{Roux2005,Franzelli2012,Galpin2008, Wang2014,See2015,Gicquel2012}.
Many of these studies applied simple kinetic mechanisms combined with the artificially thickened flame (ATF) approach \cite{Roux2005,Franzelli2012} or flamelet-based tabulated chemistry with ATF or presumed probability density function\cite{Goevert2017, Galpin2008,Wang2014,See2015}. Few studies were performed considering reduced finite-rate chemistry with the stochastic fields method as in \cite{Fredrich2019}.       
%An overview of  GT and GTMC LES can be found in~\cite{Gicquel2012}, which discusses specific issues related to GT combustion chambers, including combustion instabilities.  
%\hl{The PRECCINSTA burner has been extensively investigated with LES, proving that numerical simulations can reproduce the main experimentally observed characteristics of the flow field, both in stable and unstable regimes} e.g., in ~\cite{Roux2005, Galpin2008, Wang2014,See2015}.
%Simulations of the DLR burner showed, that the PVC is important for the flame anchoring in the dual swirl burner \cite{Chen2019}.
One fundamental aspect in numerical simulations of GTs is the modeling of heat losses for both steady \cite{Donini2016, Goevert2017, Benard2019} and oscillating operating points \cite{Kraus2017, Kraus2018}. Major heat transfer phenomena include the heat transfer from the hot product upstream, towards the swirler and the plenum, which can preheat the unburned mixture, and heat losses at the combustor walls~\cite{Kraus2017}.  
Accounting for heat transfer processes improves predictions of the temperature as well as major and minor species, not only in regions close to the walls but in the whole burner,  compared to simulations with an adiabatic wall treatment~\cite{Donini2016, Goevert2017, Benard2019,Tang2021}.
However, numerous studies demonstrated that adiabatic LES are able to reasonably predict the thermo-acoustic cycle in a combustor, see e.g. \cite{Roux2005,Franzelli2012}.   
Further studies indicated that the inclusion of heat losses increases the \mbox{prediction} accuracy of the combustion instabilities~\cite{Kraus2018,Agostinelli2021,Tay-wo-chong2017}, results in more dynamic flames~\cite{Massey2021}  and an altered flame shape due to additional flame anchoring in the outer shear layer~\cite{WoChong2012} with effects on the flame dynamics.

Recently, the SFB606 Gas Turbine Model Combustor~\cite{Arndt2015, Meier2016, Kraus2016, Arndt2017,Kraus2017} was experimentally investigated at three operating conditions, one stable (flame A) and two unstable (flame B and C). Phase-averaged PIV and Raman measurements~\cite{Arndt2015}, as well as 2-D phosphor thermometry measurements at the burner wall~\cite{Arndt2020}, were carried out. 
This burner configuration features a similar dual swirler design as the DLR dual swirl burner~\cite{Weigand2006a, Meier2006}. 
%and the SFB606 burner feature a similar dual swirler design, the two configurations differ on the fuel injection. Specifically,
However, 
%the fuel injection strategy differs between the two GTMCs: 
the DLR dual swirl burner introduces the fuel directly into the combustion chamber between the two concentric swirlers, while the SFB606 burner features fuel injection inside the swirler~\cite{Meier2006,Arndt2015}. Furthermore, the SFB606 was designed with several improvements to reduce uncertainty in the inflow boundary conditions~\cite{Arndt2015}.  
%\todo[inline]{has the old DLR combustor a similar cycle or not? i mean equivalence ration oscillation etc? So I suspect it to be similar, but I could find a detailed analysis of this cycle anywhere. Since the mixing }

Experimental observations~\cite{Arndt2015,Meier2016} indicate that the thermo-acoustic feedback mechanism of the SFB606 burner is excited by oscillations of the equivalence ratio and the convective time delay.  
Previous numerical studies of the SFB606 GTMC focused on the unstable operating point flame C ($P_{th} =$ 30 kW, $\Phi = $ 0.83, $L=$ 1.6)~\cite{Kraus2017,Kraus2018}. 
The effects of different heat transfer modeling approaches and isothermal boundary conditions on acoustic behavior were investigated using LES with a two-step kinetic mechanism and the ATF approach. The studies showed that heat losses must be taken into account to obtain an adequate prediction of the experimentally observed thermo-acoustic mode.

%In \cite{Franzelli2012, Lourier2017} the thermo-acoustic feedback mechanism of the PRECCINSTA burner was analyzed with Large Eddy Simulations~(LES), showing that because of pressure oscillations, the fuel is accumulated in the swirler during the high-pressure phase, leading to a local increase in the equivalence ratio.  This is then convected with a time delay into the combustion chamber, affecting the heat release of the flame.
%Although the thermo-acoustic feedback mechanism due to equivalence ratio oscillation and convective time delay has been qualitatively reproduced in~\cite{Franzelli2012, Lourier2017}, there has still been no detailed quantitative analysis of the flow field and flame characteristics over the oscillation phases. 

The feedback mechanism based on equivalence rate oscillations and convective time delay has been largely described in the literature, e.g., in~\cite{Lieuwen1998,Kim2010,Cosic2015}, and was reproduced for the  PRECCINSTA burner with LES~\cite{Franzelli2012, Lourier2017}. Nevertheless,  there has not yet been a detailed quantitative comparison of the flow field and flame characteristics over the oscillation phases with the experimental data. 
Therefore, the objective of this study is twofold: (1) to numerically investigate the thermo-acoustic unstable flame B ($P_{th}=\SI{25}{\kilo\watt}$,  $\Phi=\SI{0.7}{}$, $L=\SI{1.6}{}$) of the SFB606 GTMC with a flamelet-based combustion model using both adiabatic and non-adiabatic tabulated manifolds: in addition to time-averaged data, also phase-averaged data is compared to the simulations results for the first time. This allows for an additional level of detail compared to previous studies; %(2) to show the benefit in terms of predictions capabilities of LES by using an enthalpy-dependant table compared to the adiabatic table. 
\blueF (2) to show the benefit in terms of predictions capabilities of LES considering heat losses compared to adiabatic LES.\color{black}
The numerical results discussed in this work are obtained using LES combined with a flamelet progress variable combustion model and the ATF approach.
%to accurately characterize the thermo-acoustic feedback mechanism qualitatively and quantitatively; (3) to analyze not only the time-averaged  velocity and selected reactive scalars but also, for the first time, to compare the phase-averaged quantities with the experimental data.
%\color{black}
%Combustion dynamics and acoustic spectra are analyzed. \color{blue}Furthermore, phase shifts of the overall and fuel mass flux are shown.
\color{black}
 
%To better understand the complex interaction between acoustics, turbulent mixing and heat release several experimental and numerical studies on  gas turbine model combustors were conducted in the past~\cite{Weigand2006,Meier2006,Meier2007,Arndt2015, Cosic2015, Kim2010}.
%hydrodynamic instabilities and the effects of the PVC were experimentally analysed in e.g.,~\cite{Boxx2012, Stoehr2018} indicating a strong coupling between the PVC  and the thermo-acoustic cycle of the flame as well as on the flame shape. 

%% file: experimental.tex
\begin{figure}[b!]
	  \centering
	  \vspace{-20pt}
	\includegraphics[width=67mm]{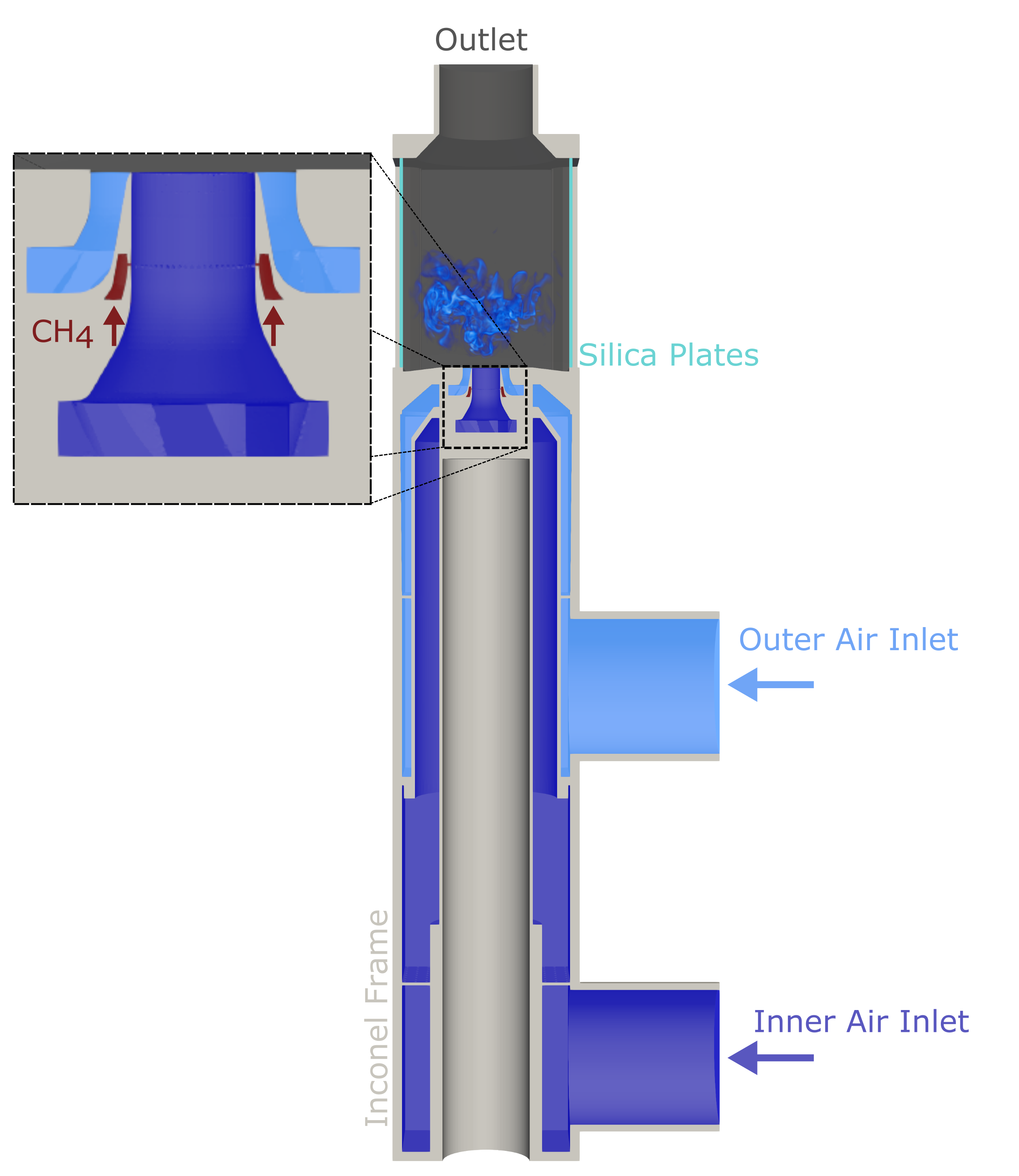}
	\vspace{-5pt}
	\caption{SCHEMATIC OF THE SFB606 GTMC BURNER SHOWING THE  DOMAIN USED FOR THE CFD SIMULATIONS.}
\label{fig:Burner_setup}
\end{figure}

\noindent In this study, the unstable operation point (flame B) of the SFB606 dual swirl GTMC is simulated and compared with \mbox{experimental} data from Arndt et al.~\cite{Arndt2015,Arndt2020}. 
A detailed description of the combustor can be found in \cite{Arndt2015, Arndt2017, Kraus2016,Kraus2017}; only a brief summary is given here. 

\color{black} 
A schematic of the burner is shown in Fig. \ref{fig:Burner_setup}.
The combustion chamber, with a square cross-section ($\SI{89}{\milli\meter}\times \SI{89}{\milli\meter}$), features a dual swirl injector where each swirler is fed by an individual air intake. Each of these intakes is equipped with a sonic nozzle (not displayed in Fig.~\ref{fig:Burner_setup}).
%removed to safe space and perforated plates to ensure that the boundary conditions are well-defined and homogeneous.
The inflow gases are injected into the combustion chamber at ambient pressure and temperature. 
 \ce{CH4} is injected into the inner swirler ($D=\SI{15}{\milli\meter}$) \SI{12}{\milli\meter} below the chamber inlet through 60 holes arranged in a circle with a diameter of  \SI{0.5}{\milli\meter}  \cite{Arndt2015,Arndt2017}. The annular outer swirler ($D_i=\SI{15.2}{\milli\meter}$, $D_o=\SI{24}{\milli\meter}$) provides air only. At the outlet, the burner is connected to the ambient environment through a circular pipe, while optical access is provided by four windows. 
\noindent Pressure oscillations were measured using microphone probes in the combustion chamber and both air inlets.
Velocities were measured with stereoscopic particle image velocimetry (PIV) in the centerplane~\cite{Arndt2015}. 
Single-shot laser Raman scattering was used to deduce the temperature and mass fractions of \ce{O2}, \ce{N2}, \ce{CH4}, \ce{H2}, \ce{CO}, \ce{CO2} and \ce{H2O}~\cite{Arndt2015}.
%, applying the procedure described in \cite{Bergmann1998} with the mixture fraction definition given by Bilger \cite{Bilger1990}. 
The inner wall temperature $T_w$ of the silica plates and the metal frame were measured by  Arndt et al.~\cite{Arndt2020} using 2D phosphor thermometry. 
% Furthermore, \ce{OH}$^*$-chemiluminescence measurements were conducted in~\cite{Arndt2015,Kraus2016}. %removed to save space

%% file: numerics.tex
%\todo[inline]{Kinetic mechanism should be added}
\noindent The LES are performed with an in-house solver integrated into the OpenFOAM framework \cite{OpenFOAM}.
%\todo[inline]{change citation}
For combustion modeling, a flamelet-based tabulated chemistry approach, using premixed manifolds is employed \cite{Vanoijen2000} to retrieve the thermo-chemical state during the simulation. \blue The solver was 
successfully validated in previous works on LES combined with tabulated chemistry\cite{Popp2015,Gierth2018,Popp2020}.
\color{black}
The flow field is described by the Favre-filtered \mbox{Navier-Stokes} equations. Additional transport equations are solved for the mixture fraction  $\widetilde{Z}$, the progress variable $\widetilde{Y}_c=\widetilde{Y}_{CO_2} + \widetilde{Y}_{H_2O}$ and the enthalpy $\widetilde{h}$ for the non-adiabatic simulation only. These variables are employed to perform the flamelet table lookup $\widetilde{\Phi}=\Phi^{table}(\widetilde{Z},\widetilde{Y}_c)$ or  $\widetilde{\Phi}=\Phi^{table}(\widetilde{Z},\widetilde{h},\widetilde{Y}_c)$ in the adiabatic/non-adiabatic LES, respectively.
The operator $\widetilde{\cdot }$ denotes the Favre-filtered nature of the solved equations. To account for heat losses in the enthalpy-dependent table, the procedure outlined in~\cite{Ketelheun2013,Steinhausen2020} is used. 
%To account for pressure-density coupling in the simulation the density $\rho$ is not read from the table. Instead the comprehensibility $\Psi$ is tabulated and the density calculated as
%\begin{equation}
%\rho^{LES}=p^{LES}\cdot\Psi^{table}.
%\end{equation}
\noindent Since the reported pressure oscillations are below \SI{1}{\kilo\pascal} \cite{Arndt2015}, the effect of pressure variations on the chemical composition and temperature is assumed  negligible~\cite{See2015,Goevert2017}. 
%, which was shown for much large pressure ranges \cite{Huo2014}.
%\todo[inline]{here we need to say how the adiabatic table is.. 1-2 sentences max}
\noindent Flamelet tables are generated with the in-house solver ULF \cite{ulf} from calculations of freely-propagating premixed flames \blue with the GRI-3.0 mechanism\cite{grimech3}. 
\blue
For the adiabatic table, flames are calculated for different mixture fractions and parameterized by $Z$ and  the normalized progress variable $Y_c^{norm}=(Y_c-Y_{c, min})/(Y_{c, max}-Y_{c, min})$. For the enthalpy-dependent table, this step is repeated for different enthalpy levels, which are normalized to $h^{norm}$ analogous to $Y_c^{norm}$. 
\color{black}
The different enthalpy levels are obtained by varying the inlet temperature between \SI{300}{\kelvin} and \SI{900}{\kelvin}, and by a variable ratio of cooled exhaust gases, which are recirculated and mixed with the unburned mixture at \SI{300}{\kelvin}  in the flamelet calculation. The procedure is explained in detail in~\cite{Steinhausen2020}. 
The  tables contain 444$\times$101($\times$101 points) in the  $Z$, $Y_c^{norm}$ and $h^{norm}$ space. Unity Lewis number  is assumed for all species.

%In \cref{fig:table} the tabulated temperature $T$, progress variable source term $\dot{\omega}_{Y_c}$  and $OH$ mass fraction $Y_{OH}$ are shown  as function of the normalized progress variable  $Y_c^{norm}$ and enthalpy $h^{norm}$. For the mixture fraction $Z$ stochiometric conditions are assumed. 
%\begin{figure*}[h]
%	\centering
%	\includegraphics[width=1.0\textwidth]{parts/setup/table_fields.png}
%	\caption{Temperature $T$ , progress variable source term $\dot{\omega}_{Y_c}$ and $OH$ mass fraction $Y_{OH}$ from the table plotted over $Y_c^{norm}$ and $h^{norm}$ for a stochiometric mixture fraction.}\label{fig:table}
%\end{figure*}
\noindent The effects of unresolved turbulence are accounted for by the $\sigma$ sub-grid model \cite{Nicoud2011} with $c_\sigma=1.5$.  
The turbulence-chemistry interaction is modeled using the ATF approach, following the procedure presented in~\cite{Colin2000}. In the ATF approach, the flame front is artificially thickened using the thickening factor $\mathcal{F}$, enabling it to be resolved on coarse grids, while preserving the flame propagation. The effect of unresolved flame wrinkling is accounted for by the efficiency function $\mathcal{E}$. Here, the efficiency function formulation developed by Charlette et al.~\cite{Charlette2002} is used. 
To keep regions of pure mixing unaltered and only thickening the regions where combustion takes place the flame sensor formulated in~\cite{Popp2019} is used. 
A grid-adaptive thickening with six grid points in the flame zone is applied following~\cite{Charlette2002}. 
%The maximum thickening factor $\mathcal{F}_{max}$  is defined by the number of points in the flame zone, set here to six, and the tabulated laminar flame thickness. 

%Thus, the flame thickening $\mathcal{F}$ is calculated as
%\begin{align}
%\mathcal{F}&=1+\Omega(\mathcal{F}_{max}-1) \quad \text{with}\\
%\Omega&=\dfrac{\nabla \PV}{(\nabla \PV)_{max}}+\dfrac{\dot{\omega}_{\PV}}{(\dot{\omega}_{\PV})_{max}} \left(1-\dfrac{\nabla \PV}{(\nabla \PV)_{max}}\right),
%\end{align}
%where $\Omega$ denotes the flame sensor,  constructed with the normalized progress variable gradient  and the normalized progress variable source term.  The maximum thickening factor $\mathcal{F}_{max}$  is defined by the number of points in the flame zone, set here to six, and the tabulated laminar flame thickness \cite{Popp2019}. 
\noindent To account for heat conduction through the solid combustion chamber and the swirler, a loosely coupled simulation of the solid and fluid domain is performed. 
Calculations of the solid domain were carried out with the OpenFOAM solver chtMultiRegionFoam (v2006), where the enthalpy equation is solved in the solid. Previous studies ~\cite{Kraus2018} showed, that the solid domain is in a steady state as the heat release fluctuations in the fluid are too fast to alter the temperature distribution in the solid. Therefore, the solid domain is simulated separately from the fluid domain, drastically reducing the computational cost. To reach equilibrium between the solid and the time-averaged fluid domain 4 iterations of 30~ms of fluid simulation and a consecutive solid simulation with the fluid heat flux as boundary condition were performed. Further iterations did not alter the resulting temperature field in the solid.
%\begin{equation}
%\frac{\partial(\rho h)}{\partial t}=\frac{\partial}{\partial x_{j}}\left(\alpha \frac{\partial h}{\partial x_{j}}\right),
%\end{equation}
%is solved, where $\alpha$ is the thermal diffusivity, $\rho$ the density and $h$ the enthalpy in the solid. \\\\

%\include{asme_paper_revision/parts/apriori}

%% file: numerical_setup.tex
\noindent The numerical domain contains both air inlet sections and the combustion chamber as shown in Fig.~\ref{fig:Burner_setup}. The grid contains 17 million cells (mostly hexahedra). 
%A section of the mesh is shown in Fig.~\ref{fig:mesh}. 
A slice of the mesh is shown on top of the time-averaged temperature $\langle \tilde{T}\rangle$ field in Fig.~\ref{fig:mesh}. 
At the inlets, air and methane are injected into the domain at \SI{300}{\kelvin}. The mass flows are $\dot{m}_{air,i}=\SI{282}{\gram\per\min}$, $\dot{m}_{air,o}=\SI{451}{\gram\per\min}$ and \mbox{$\dot{m}_{CH_4}=\SI{30}{\gram\per\min}$} for the inner, outer air and methane inlets, respectively~\cite{Arndt2015}.
\blue
The pressure $\overline{p}_{far}=\SI{1}{atm}$ is prescribed at the outlet with the approximate waveTransmissive boundary condition\cite{OpenFOAM}, which models 
the impedance of an open duct.
\blueF LES with adiabatic walls are performed using the adiabatic table, 
  while the simulation with isothermal boundary conditions at the chamber walls requires the enthalpy-dependent table. \color{black}
The wall temperature $T_w$ on the silica plates and the chamber corners are interpolated from the phosphor thermometry measurements from \cite{Arndt2020}, while the temperature at the burner plate and the swirler wall is determined by CHT calculation. 

\noindent The solid domain is discretized with 13 million cells. 
In the CHT calculation, the heat conduction is solved in the solid parts of the burner, namely the silica plates and the Inconel frame. Where available, wall temperatures are interpolated from measurements \cite{Arndt2020}, while the time-averaged heat flux from the fluid domain is used as a boundary condition for the inner wall.  At the outer walls of the solid, a heat transfer coefficients of $h=\SI{112}{\watt\per\square\meter\per\kelvin}$ and an ambient temperature of $T_{amb}=\SI{300}{\kelvin}$ are set. Both are estimated from the reported heat loss in \cite{Arndt2020}.
% The heat transfer coefficient $h$ is calculated from $\dot{Q}=A\, h\,\left(T_{s}-T_{amb}\right)$, using  the heat loss $\dot{Q}=\SI{2.2}{\kilo\watt}$ and average wall temperature $T_s\approx900K$ at the outer glass wall for flame A taken from \cite{Arndt2020}. 
\noindent The resulting steady-state $T_w$ distribution from the solid simulation is mapped onto the fluid simulation and employed as a temperature boundary condition.  This procedure was repeated until the temperature distribution in the solid reached a steady state. Afterwards the sampling was conducted over \SI{80}{ms}.
Simulations are solved using second-order schemes for space and time discretization. 
%The coordinate system used in the following has its origin in the center of the swirler at the burner exit plane and is shown in Fig.~\ref{fig:mesh}. 

%% file: results.tex
\begin{figure}[h]
	%\centering
	\includegraphics[width=90mm, trim={0 1cm 0 2.5cm},clip]{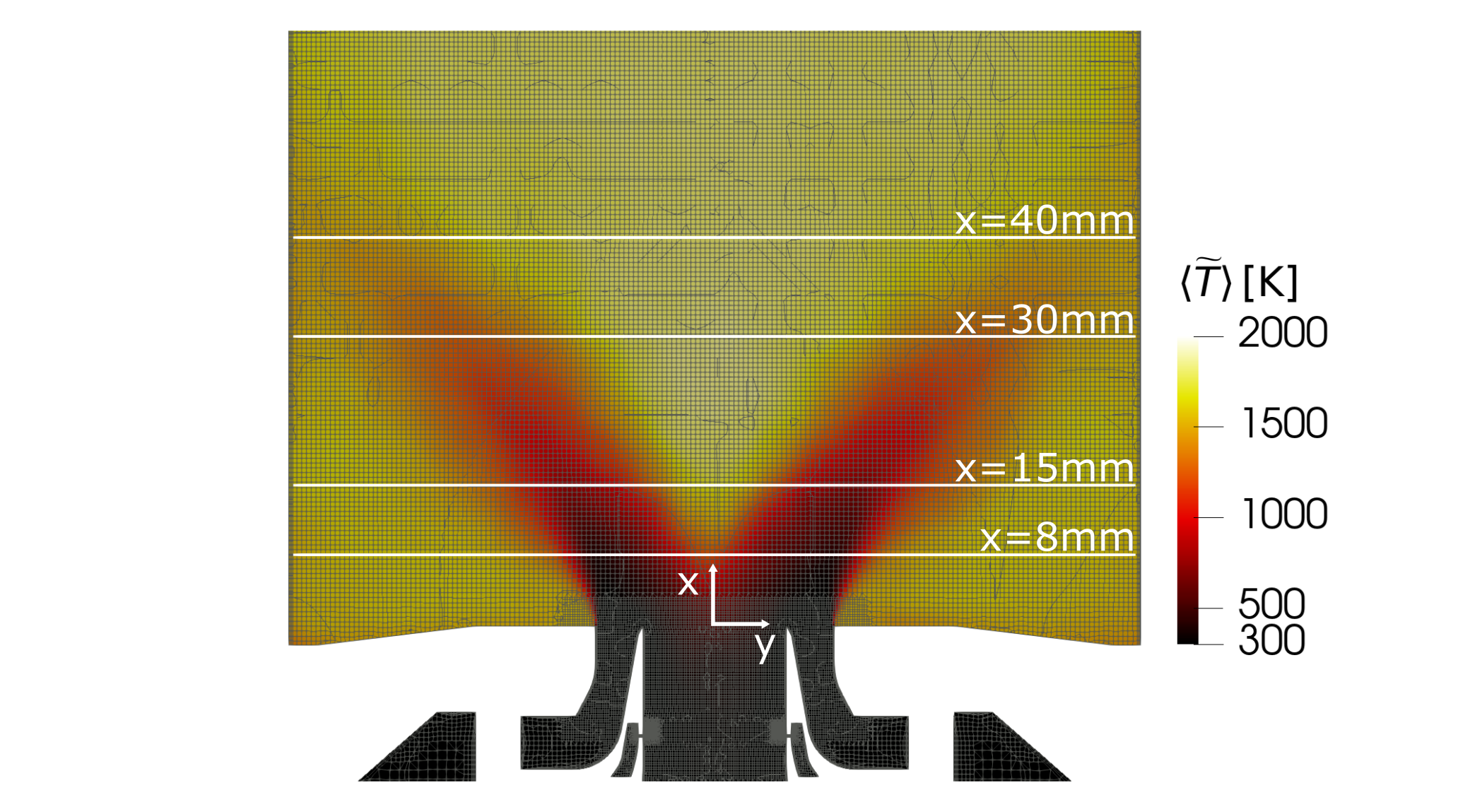}
	\vspace{-10pt}
	\caption{SLICE OF THE TIME-AVERAGED TEMPERATURE FIELD $\langle \tilde{T}\rangle$ OF THE ENTH. DEPENDENT LES. ALSO SHOWN ARE THE COORDINATE SYSTEM, THE GRID AND AXIAL POSITIONS OF QUANTITATIVE COMPARISONS.}\label{fig:mesh}
	\vspace{-10pt}
\end{figure}

\noindent In this section, the time-averaged flow field quantities are first analyzed for \blue both the adiabatic (Adi.) and enthalpy-dependent LES (Enth.). \blueF The latter is characterized by isothermal boundary conditions, requiring an enthalpy-depended table. \color{black} Subsequently, the phase-averaged variables and acoustic cycle obtained in the enthalpy-dependent LES are discussed. 
The time-averaged temperature $\langle \tilde{T}\rangle$ field of the enthalpy-dependent LES is plotted in Fig.~\ref{fig:mesh}. Also shown are the mesh and the coordinate system, which has its origin in the center of the swirler at the burner exit plane. The white lines mark the axial positions at which quantitative comparisons are presented in the following.

\subsection{Time-averaged flow field}
\begin{figure}[h]
	\centering
	\includegraphics{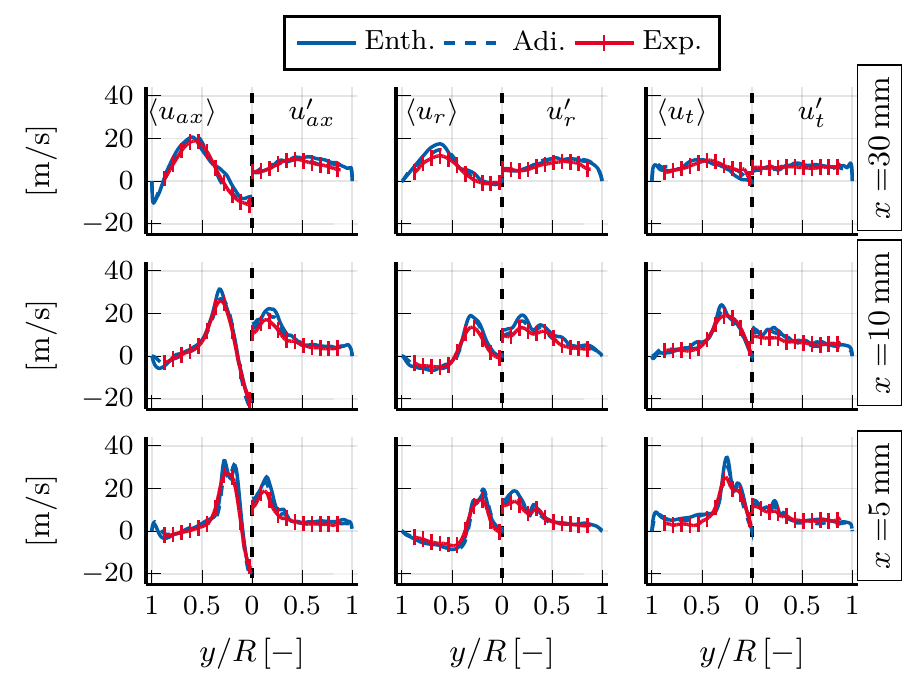}
	\caption{COMPARISON OF TIME-AVERAGED (LEFT) AND RMS (RIGHT) VELOCITY PROFILES BETWEEN SIMULATIONS AND EXPERIMENTAL DATA AT FOUR AXIAL LOCATIONS ($x=\SI{5}{\milli\meter}$, $x=\SI{10}{\milli\meter}$, $x=\SI{20}{\milli\meter}$ AND $x=\SI{30}{\milli\meter}$). FROM LEFT TO RIGHT: AXIAL $u_{ax}$, RADIAL $u_r$ AND TANGENTIAL $u_t$ VELOCITY COMPONENTS. }\label{fig:meanPIV}
	
\end{figure}

\begin{figure}[h]
	\centering
	\includegraphics{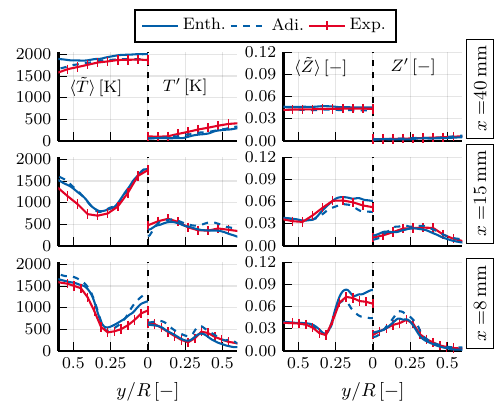}
	\vspace{-10pt}
	\caption{COMPARISON OF TIME-AVERAGED AND RMS TEMPERATURE (LEFT) AND MIXTURE FRACTION (RIGHT) PROFILES BETWEEN SIMULATION AND EXPERIMENTAL DATA AT THREE AXIAL LOCATIONS ($x=\SI{8}{\milli\meter}$, $x=\SI{15}{\milli\meter}$ AND $x=\SI{40}{\milli\meter}$). }
	\label{fig:meanTZ}
\end{figure}

\begin{figure}[hb!]
	\centering
	\includegraphics{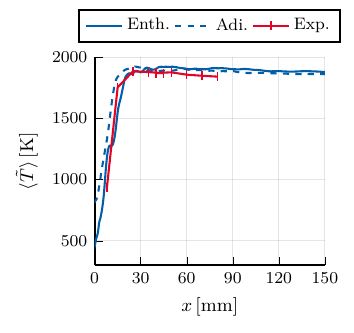}
	\vspace{-12pt}
	\caption{COMPARISON OF TIME-AVERAGED TEMPERATURE PROFILES BETWEEN SIMULATIONS AND EXPERIMENTAL DATA ALONG  THE CENTERLINE OF THE COMBUSTOR.}
	\label{fig:centerlineT}
\end{figure}
\noindent In Fig.~\ref{fig:meanPIV}, the time-averaged velocity profiles obtained by the LES and the experimental measurements are compared at three axial locations (left half of the plots). Good agreement is observed for all velocity components. 
The position and the value of the velocity peaks are correctly predicted at all axial locations. The simulations accurately reproduce the inner recirculation zone (IRZ) indicated by the negative velocities at $y/R=0$, and the outer recirculation zones (ORZ), marked by negative velocities at $y/R=\pm1$. The RMS profiles of the velocity are also shown in Fig.~\ref{fig:meanPIV} (right half of the plots). Overall a very good prediction is observed \blue for both adiabatic and enthalpy-dependent~LES. \color{black}
%Only peak values are slightly overestimated by the LES indicating that the main features of the fluctuating flow field are captured. 

Figure \ref{fig:meanTZ} shows the time-averaged temperature $\langle\widetilde{T}\rangle$ and \mbox{mixture} fraction profile $\langle\widetilde{Z}\rangle$ at three axial positions (left) as well as the RMS values of both variables (right). %Due to the limited optical access, Raman measurements are only available for $y/R<0.6$.
\blue
The adiabatic LES overestimates the temperature in the ORZ regions since heat losses are not accounted for. Furthermore, the flame anchoring is predicted much closer to the inlet as indicated also in the mean temperature profile along the centerline shown in Fig.~\ref{fig:centerlineT}. At $x=\SI{8}{\milli\meter}$ $\langle\tilde{Z}\rangle$ is underpredicted by the adiabatic LES in the center, indicating that the mixture field is not captured accurately. 
Regarding the enthalpy-dependent LES, it can be seen that at $x=\SI{8}{\milli\meter}$ the temperature at the center of the burner is slightly overestimated due to the prediction of the hot IRZ slightly closer to the chamber inlet. 
This has also been observed in Fig.~\ref{fig:centerlineT}, which indicates that the profile shifts slightly upstream, towards the inlet, for the enthalpy-dependent LES  \color{black}, although the general shape is well captured.

\begin{figure}[t]
	\centering
	\includegraphics{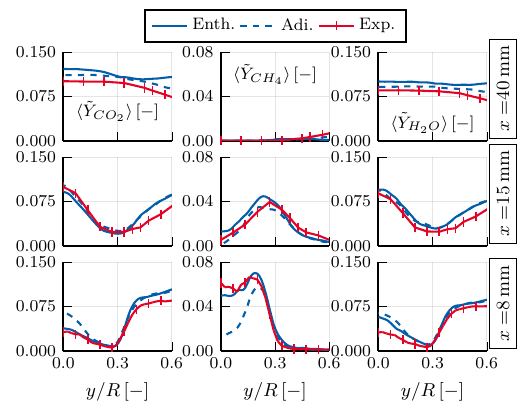}
	\vspace{-10pt}
	\caption{COMPARISON OF TIME-AVERAGED SPECIES MASS FRACTION PROFILES BETWEEN SIMULATION AND EXPERIMENTAL DATA AT THREE AXIAL LOCATIONS ($x=\SI{8}{\milli\meter}$, $x=\SI{15}{\milli\meter}$ AND $x=\SI{40}{\milli\meter}$).}\label{fig:meanSpecies1}
	\vspace{-15pt}
\end{figure}

\noindent Figure~\ref{fig:meanSpecies1} shows the time-averaged profiles of  \ce{CO2}, \ce{CH4}, and \ce{H2O} mass fractions. 
\blue At $x=\SI{8}{\milli\meter}$, \ce{CH4} is severely underpredicted by the adiabatic LES. This is partially caused by an altered mixing field,  which transports less fuel to the center of the combustor, indicated by the $\langle\tilde{Z}\rangle$ profile, see Fig.~\ref{fig:meanTZ}, and by the flame anchoring much closer to the inlet plane, see Fig.~\ref{fig:centerlineT}, already consuming \ce{CH4}.
Downstream both simulation results are close to the experimental measurements. The findings showcase that the adiabatic simulation is not able to capture the correct flame anchoring and mixture field, therefore it is not considered for the analysis of the acoustic cycle in the following.
\color{black}
%The simulation results closely reproduce the measured results, similarly to the temperature and mixture fraction profiles in Fig.~\ref{fig:meanTZ}. At $x=\SI{40}{\milli\meter}$ both   \ce{CO2} and \ce{H2O} are overestimated, while \ce{CH4} is underestimated, indicating, that the simulation consumes the fuel slightly faster as the experiments suggest. At $x=\SI{8}{\milli\meter}$,  \ce{H2O} is overestimated in the center, while \ce{CH4} is underestimated directly in the center. This is again caused by the slight shift in the flame anchoring point as discussed in Fig.~\ref{fig:centerlineT}.

\subsection{Analysis of the acoustic cycle}
 As discussed in the previous section, the time-averaged results over the entire thermo-acoustic cycle indicate good agreement between the enthalpy-dependent LES and experiments. 
In this section, the pressure response of the burner is first analyzed, then a detailed analysis  of the  flow field over the different phases in the cycle is conducted.
%\blue Here only the enthalpy dependent LES, is considered, since the adiabatic simulation proved to be unable to reproduce the correct shape and behaviour. \color{black}

\begin{figure}[t!]
	\centering
	\includegraphics{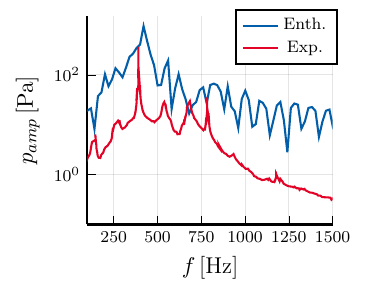}
	\vspace{-10pt}
	\caption{CALCULATED AND MEASURED PRESSURE SPECTRA INSIDE THE COMBUSTION CHAMBER.}
	\label{fig:frequency_responce}
\end{figure}

The Fourier transformed pressure amplitude in the combustion chamber $p_c$ calculated in the LES is plotted in Fig.~\ref{fig:frequency_responce} and compared with the measured pressure signal. 
%The dominant frequency of \SI{392}{\hertz} is slightly overestimated, by \SI{30}{\hertz}, while the amplitude of the dominant mode is overestimated by \SI{400}{Pa}. 
\blue The dominant frequency of \SI{392}{\hertz} corresponding to the resonance mode of the inner air intake section \cite{Arndt2015}, is captured closely in the simulation.  \color{black}
The other peaks in the experimental spectrum, corresponding to the second harmonic of the dominant mode (\SI{785}{\hertz}) and the Helmholtz frequency of the combustor and outer plenum (\SI{139}{\hertz})\cite{Arndt2015}, are not captured by the simulation. 
%\hl{The general amplitude level is overestimated by the simulation.}
\blue
%Lourier et al.~\cite{Lourier2017} \hl{showed that the glass walls of the PRECCINSTA burner strongly dampen pressure amplitudes. This is true not only for the dominant mode but also for other frequencies. The overestimated pressure levels could therefore be partly caused by the neglected wall dampening. Another cause might be the outlet boundary condition being too reflective.}
%\hl{The overestimated amplitudes could be caused by uncertainties in the boundary conditions.} 
Previous studies on a similar burner configuration suggested that the side glass walls could significantly dampen pressure amplitudes~\cite{Lourier2017}, \mbox{depending} on their mounting design.

\color{black}
%\todo[inline]{this part needs to be adjusted}

\begin{figure*}[h]
	\centering
	
	\includegraphics[width=150mm , trim={0 0 0 0.7cm},clip]{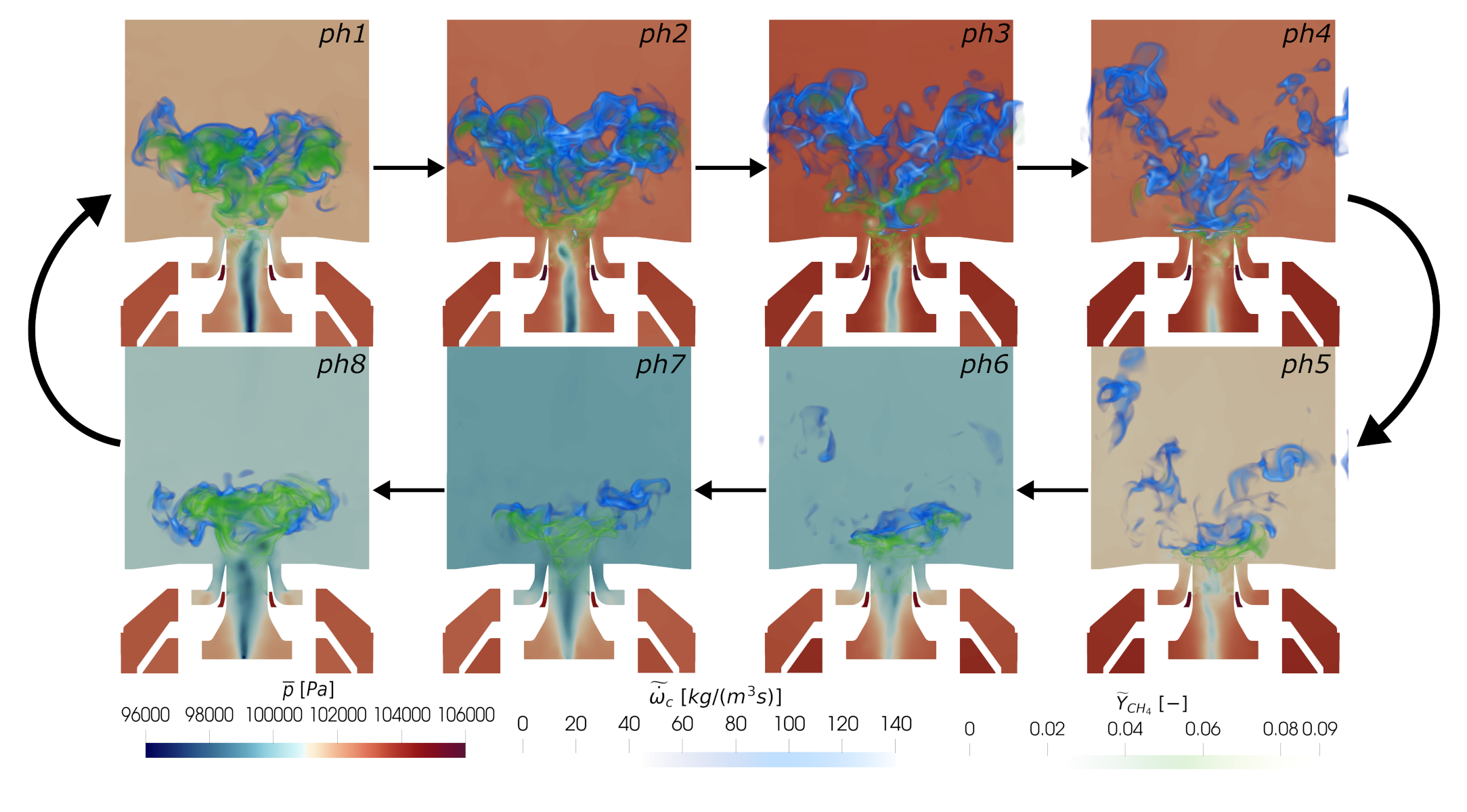}
	\vspace{-12pt}
	\caption{VISUALIZATION OF THE FEEDBACK CYCLE WITH VOLUME PLOTS OF METHANE MASS FRACTION $\widetilde{Y}_{CH_4}$ AND PROGRESS VARIABLE SOURCE TERM  $\widetilde{\dot{\omega}}_c$. THE COLORED AREA IN THE CENTRAL PLANE INDICATES THE PRESSURE $\overline{p}$.}\label{fig:cycle}
\end{figure*} 
\blue Experimental investigations  in~\cite{Arndt2015, Meier2007} showed that equivalence ratio oscillation coupled with a convective time delay contribute to the thermoacoustic feedback mechanisms.
\color{black}

The LES reproduces the same behavior. To analyze the feedback mechanism in detail, a single cycle is shown in Fig.~\ref{fig:cycle},  where eight snapshots
are arranged from a phase angle of $0^\circ$ (\textit{ph1}) for the sin-pressure signal to a phase angle of $337.5^\circ$ (\textit{ph8}), similarly to the experimental measurements from~\cite{Arndt2015}.
Each snapshot features the instantaneous three-dimensional progress variable source term, the mass fraction of methane and the pressure contour on the center plane.
%\todo{phase angle 1 is from 0 or -22.5?}
 
\begin{figure}[b!]
	\centering
	\includegraphics{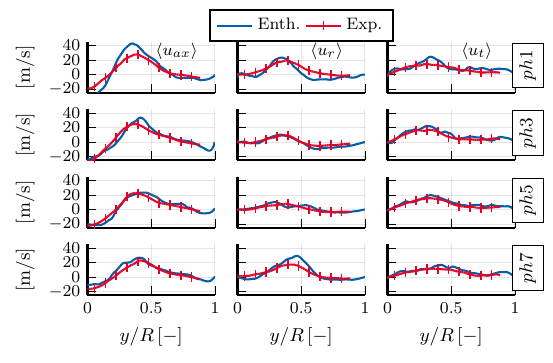}
	\vspace{-20pt}
	\caption{COMPARISON OF PHASE-AVERAGED VELOCITY PROFILES AT~$x=\SI{15}{\milli\meter}$. FROM LEFT TO RIGHT AXIAL: $u_{ax}$, RADIAL $u_r$ AND TANGENTIAL $u_t$ VELOCITY COMPONENTS.}
	\label{fig:phasePIV}
	\vspace{-10pt}
\end{figure}

 For quantitative analysis, the phase-conditioned averaged velocity field has been calculated for the eight phases following the procedure outlined in~\cite{Arndt2015}, i.e. samples are sorted and averaged in a $45^\circ$ phase angle range around the previously defined phase centers and compared with the phase-conditioned averaged experimental data in Fig.~\ref{fig:phasePIV}. The symmetry of the combustion chamber was utilized to increase the number of samples, gathered over  \SI{31}{\milli\second} of simulated time. 
%LES results were collected over 13 cycles, corresponding to \SI{31}{\milli\second}.
%To quantitative information the velocity profiles at an representative position , $x=\SI{15}{\milli\meter}$, are phase-averaged  following the procedure outlined in~\cite{Arndt2015}. Todo so all samples in a $45^\circ$ range around the previously defined phase centers are averaged and shown in Fig.~\ref{fig:phasePIV}.
\noindent 
A PVC structure can be observed in the low-pressure region inside the swirler over all phases. 
\textit{ph3} corresponds to the phase with the highest chamber pressure, while \textit{ph7} features the lowest pressure.  
In \textit{ph3}  the high pressure is linked to a large flame zone which has a visible V-shape in Fig.~\ref{fig:cycle}.
Due to the high chamber pressure, the driving pressure difference between the plenum and the chamber $\Delta P$ is low and the velocity inside the swirler and plenum \mbox{decreases,} leading to the accumulation of fuel inside the swirler close to the inlet, slimilarly to~\cite{Meier2007, Lourier2017}.
\noindent In \textit{ph4} and \textit{ph5}, the flame zone is convected downstream until the fuel is completely consumed. In these phases the flame interacts with the chamber walls, see Fig.~\ref{fig:cycle}.
Up to \textit{ph7}, the chamber \mbox{pressure} drops and the flame zone is in the center of the burner close to the inlet, where the fresh mixture accumulates. At \textit{ph7} the pressure difference between chamber and plenum is largest, accelerating the gas inside the swirler, which in turn convects the fuel-rich mixture into the chamber, where a larger flame zone forms in the subsequent phases. The thermal expansion of the gas then leads to an increase in chamber pressure and a consequent decrease in the pressure difference $\Delta P$. 

\noindent The phase-averaged velocity components shown in  Fig.~\ref{fig:phasePIV} indicate that at \textit{ph7}  the \mbox{simulation} features higher radial velocities compared to the experiments at the reference location $x=\SI{15}{\milli\meter}$. 
Axial velocity peaks are overestimated until \textit{ph3}. 
Here, the  LES  predicts the measured velocity accurately. 
This indicates that the stronger pressure oscillations predicted by the simulation,  yield higher inflow velocities in the phases of low but rising pressures.
%The flame response over the cycle is evaluated by analyzing the phase-averaged temperature $T$ and \ce{CO2}  mass fraction profiles at the axial position $x=\SI{15}{\milli\meter}$. Both curves exhibit good agreement from \textit{ph7} to \textit{ph1} (see Fig.~\ref{fig:phaseRaman}), while deviations are observed in the other phases. 
%The agreement is then opposite in phase to that observed for the velocity, suggesting that the phase-dependent errors in the velocity are coupled with deviations in flame prediction in the subsequent phases. 
%\begin{figure}[h]
%	\centering
%	\includegraphics{../op2-plots/TZ_phases.pdf}
%	\caption{COMPARISON OF PROFILES FOR THE PHASE-AVERAGED TEMPERATURE $T$ (LEFT) AND \ce{CO2} MASS FRACTION $Y_{CO2}$ (RIGHT) AT $x=\SI{15}{\milli\meter}$.}
%	\label{fig:phaseRaman}
%\end{figure}
%This is consistent with the overestimation of the mixture fraction over the phases. 

\begin{figure}[t]
	\centering
	\vspace{-5pt}
	\includegraphics{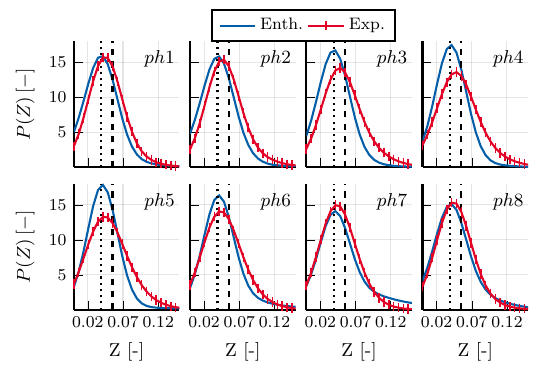}
	\vspace{-15pt}
	\caption{PHASE-AVERAGED MIXTURE FRACTION PDF  AT $x=\SI{15}{\milli\meter}$.  DASHED AND DOTTED LINES MARK THE STOICHIOMETRIC AND GLOBAL MIXTURE FRACTION, RESPECTIVELY.}
	\label{fig:phaseZ}
	\vspace{-5pt}
\end{figure}

\begin{figure}[h]
	\centering
	\vspace{-5pt}
	\includegraphics{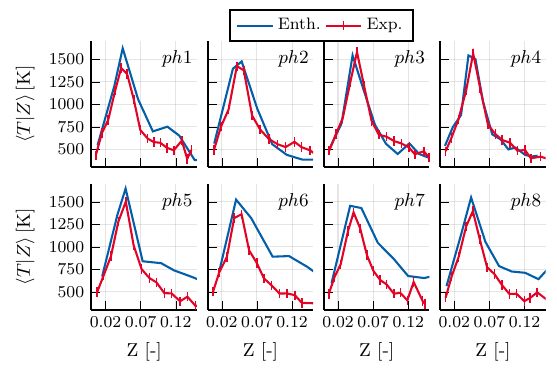}
	\vspace{-20pt}
	\caption{TEMPERATURE CONDITIONED ON MIXTURE FRACTION OVER THE DIFFERENT PHASES AT $x=\SI{15}{\milli\meter}$. }
	\vspace{-5pt}
	\label{fig:phaseTZ}
\end{figure}

\noindent Figure~\ref{fig:phaseZ} shows the mixture fraction PDF $P(Z)$ over the phases at the axial location $x=\SI{15}{\milli\meter}$ compared with the measurements. 
\blue
From \textit{ph3} to \textit{ph6}, the LES  predicts a slightly different profile of the mixture fraction PDF,  overestimating the peak value and underestimating the peak position. 
%and under-predicts from the LES is overestimates 
%the experimental mixture fraction PDF is  broadens towards higher mixture fraction values, while the $P(Z)$ peaks of the simulation remains at the global mixture fraction.
%From \textit{ph3} to \textit{ph6}, the experimental mixture fraction PDF broadens towards higher mixture fraction values, while the $P(Z)$ peaks of the simulation remains at the global mixture fraction.
 \color{black}
%\hl{The differences indicate that in the simulation the thermo-acoustic mode is less affected by the phases, while the experiment also features some broadening of mixture fraction PDF.}
Figure~\ref{fig:phaseTZ} shows the conditional mean of the temperature with respect to the mixture fraction.  Here good agreement with the experimental data is observed except for \textit{ph5} to \textit{ph7}, where the temperature is overpredicted for high mixture fraction values. Contrary to the experiment, pockets of high mixture fraction already enter the flame zone or the hot recirculation zone, which does not occur in the experiment.
% This might also explains the overestimated \ce{CO} mass fractions in Fig.~\ref{fig:meanSpecies1}.
The deviations between the LES and experiments in the mixture fraction-conditioned temperature are shifted compared to the deviations in $P(Z)$  by two phases.
%These results indicate that although the frequency of the oscillations is reproduced well in the simulation,  the overestimated amplitude of the oscillation does change the detailed behavior of the flame. 
These results indicate that although the frequency of the oscillation is reproduced well in the simulation, slight differences in the flame behavior are still observable due to the overestimated amplitude of the oscillation.
\blue Further investigations to address this issue will be scope of future studies.
 \color{black}

\begin{figure}[b]
	\centering
	\vspace{-5pt}
	\includegraphics[trim=0.2cm 0cm 0cm 0cm,clip]{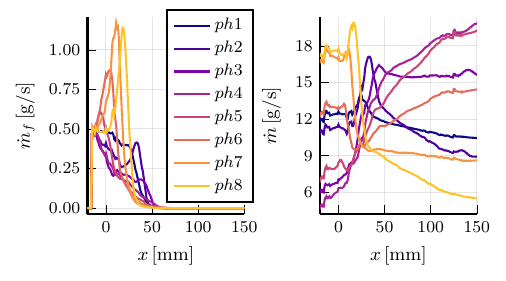}
	\vspace{-5pt}
	\caption{CROSS SECTION AND PHASE-AVERAGED PROFILES OF  FUEL MASS FLOW $\dot{m}_f$ AND GLOBAL MASS FLOW $\dot{m}$.}\label{fig:1D}
	\vspace{-12pt}
\end{figure}

\noindent 

\begin{figure}[t]
	\centering
	\includegraphics{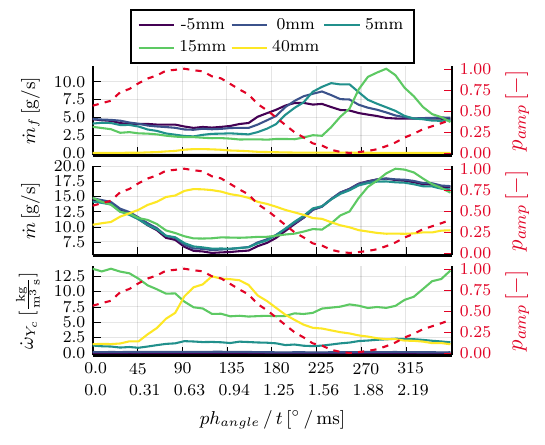}
	\vspace{-15pt}
	\caption{CROSS-SECTION AND PHASE-AVERAGED PROFILES OF  FUEL MASS FLOW $\dot{m}_f$, GLOBAL MASS FLOW $\dot{m}$ AND PROGRESS VARIABLE SOURCE TERM $\dot{\omega}_{Y_c}$ OVER THE PHASE ANGLE AT DIFFERENT AXIAL POSITIONS. THE PRESSURE SIGNAL IS INDICATED BY THE DASHED LINE. }
	\label{fig:1D_angle}
	\vspace{-5pt}
\end{figure}

To better understand the nature of the acoustic coupling to convective time delay, the fuel mass flow $\dot{m}_f$ and the global mass flow $\dot{m}$ averaged over cross-sections and phases are plotted in Fig. \ref{fig:1D}. 
The  $\dot{m}_f$ phase profiles indicate that the fuel injected into the burner (injection holes at $x=\SI{-12}{\milli\meter}$) is constant over all phases, supporting the hypothesis in~\cite{Meier2007, Arndt2015} that the fuel stream at the injection holes is not influenced by the acoustic cycle due to the high-pressure loss experienced during the injection process.
\ce{CH4} accumulates downstream and is convected in larger quantities through the chamber in the low-pressure phase (\textit{ph7}) while it is consumed in the high-pressure phase (\textit{ph3}).
\blue
Up to $x=\SI{10}{\milli\meter}$ the overall mass flow $\dot{m}$ on the contrary does not change with $x$ for all phases. For $x>\SI{10}{\milli\meter}$ a distinct peak is observed between \textit{ph8} and \textit{ph2} only.
\color{black}
%Furthermore, the peak positions of $\dot{m}_f$ coincide with that of $\dot{m}$  for all phases. 
%These findings indicate that both the overall mass flow and the fuel mass flow experience a convective delay. 
To further analyze the convective delay, the mass flows and the progress variable source term are plotted over the phase angle in Fig.~\ref{fig:1D_angle} for different axial positions. The pressure fluctuation is also shown.
\blue
The evolution of $\dot{m}_f$ shows a shift in the peak towards higher phase angles with increasing axial position, while $\dot{m}$ features overlapping sinus signals from $x=\SI{-5}{\milli\meter}$ to $x=\SI{5}{\milli\meter}$, 40$^\circ$ behind $p$, while a convective delay is observed for $\dot{m}$ for $x>\SI{10}{\milli\meter}$.
This confirms that the fuel flow experiences a convective delay already inside the swirler. 
The convective time delay of $\dot{m}_f$ between $x=\SI{-5}{\milli\meter}$ and $x=\SI{0}{\milli\meter}$ corresponds to $\Delta t= \Delta x / u_{ax}=\SI{0.17} {\milli\second}$, close to the observed phase delay of 25$^\circ$(\SI{0.174}{\milli\second}).
Between $x=\SI{-5}{\milli\meter}$ and $x=\SI{15}{\milli\meter}$ the shift between $\dot{m}_f$ and the $\dot{m}$ peak shrinks from 90$^\circ$ to only 18$^\circ$. %(\SI{0.625}{\milli\second})   ~(\SI{0.125}{\milli\second})
At $x=\SI{40}{\milli\meter}$ both are in phase of each other having experienced a similar convective delay  from  $x=\SI{15}{\milli\meter}$. % of 160$^\circ$~(\SI{1.11}{\milli\second}) and 178$^\circ$~(\SI{1.235}{\milli\second}), since $x=\SI{15}{\milli\meter}$ for $\dot{m}$ and $\dot{m}_f$, respectively. 
%Between $x=\SI{-5}{\milli\meter}$ and $x=\SI{40}{\milli\meter}$ $\dot{m}_f$ experiences a delay of 260$^\circ$(\SI{1.81}{\milli\second}), while for $\dot{m}$ only 165$^\circ$(\SI{1.146}{\milli\second}) are observed. 
The progress variable source term $\dot{\omega}_{Y_c}$ in Fig.~\ref{fig:1D_angle} showcases a high reaction rate over the entire cycle, with a broad peak around 0$^\circ$,  60$^\circ$(\SI{0.416}{\milli\second}) earlier than the  $\dot{m}_f$ peak for $x=\SI{15}{\milli\meter}$. 
The peak corresponds to the compact flame in \textit{ph1} (Fig.~\ref{fig:cycle}).
For $x=\SI{40}{\milli\meter}$ $\dot{\omega}_{Y_c}$ is limited to a broad peak around 135$^\circ$. 
\color{black}

%This confirms that the convective delay in the fuel flow already starts inside the swirler and makes it necessary to simulate the swirler and the mixing region.
%In contrast, the overall mass flux evolves according to the acoustic eigenmode of the intake section. 
%It is observed that the $\dot{m}$ signal in the swirler is delayed by about $210^\circ$ (\SI{1.41}{\milli\second}) compared to the pressure signal. 
%\todo[inline]{here we can write that delay in sec}
%\noindent  Downstream a convective delay modifies the phase shift to about $30^\circ$ (\SI{0.2}{\milli\second}), as can be seen at  $x=\SI{40}{\milli\meter}$. 
%Lastly, the progress variable source term reaches its highest amplitude at  $x=\SI{40}{\milli\meter}$, where the main flame zone is located, similarly to the experimental observations~\cite{Arndt2015}.  At this position, the progress variable source term signal evolves in phase with the pressure. \hl{ From this plot, the differences between the mass flow and mixture fraction oscillations can be drawn.}

%\FloatBarrier

%% file: conclusion.tex
This study numerically investigated the thermo-acoustic unstable operating point B of the SFB606 GTMC using LES. \blue
In one simulation adiabatic walls were assumed, while in a second LES  isothermal walls with temperatures interpolated from measurements and a coupled CHT simulation to account for the heat losses were considered. Both simulations employed a correspondent tabulated chemistry manifold coupled with the ATF  model for the turbulent combustion closure. 
\color{black}
The following conclusions can be drawn based on this study:
\begin{enumerate}
	%\item A \textit{prior} analysis of adiabatic and enthalpy dependent \mbox{tables} on the measurement data revealed that heat losses must be accounted for to capture  the cause and effect chain of the thermo-acoustic cycle correcly. Already for time-averaged results, the adiabatic table overstimates the temperature field. Furthermore, the shift in the predicted temperature over the different phases indicates that the heat losses experienced by the flow field vary over the acoustic cycle, leading to altered $P(T)$ shapes for the adiabatic table compared to the experiment. Additionally, the $P(T)$ profiles imply that the fresh mixture is preheated in the swirler.
	
\blue
	\item For the adiabatic LES a flame anchoring point much closer to the inlet plane and an altered mixture field was observed. Therefore, further phase-conditioned analysis with the adiabatic LES are not conducted.

	\item The enthalpy-dependent LES is able to closely predict the time-averaged profiles and frequency of the dominant mode, validating the employed numerical methodology as a viable tool to
	simulate thermo-acoustic oscillations and capture the feedback mechanism qualitatively.
	\blueF An equivalence ratio oscillation coupled to the dominant mode through a convective time delay, which is different from the delay of the overall mass flux, is observed and contributes to the feedback mechanism of the thermo-acoustic coupling.
	%It was confirmed that the feedback mechanism of thermo-acoustic oscillations is strongly influenced by the acoustic coupling of equivalence ratio fluctuations and a convective time delay to the resonance mode of the inner air intake section.
	\color{black} %In line with the experimental observation, 
	\item For the first time,  the phase-conditioned velocities, mixture fraction PDF and conditional temperature, have been compared with phase-conditioned PIV and Raman measurements for the enthalpy-dependent simulation. The presented numerical framework showcases fair quantitative agreement with the phase-averaged measurements, although small deviations in the pressure amplitude and phase-dependent mixing field are observed. 
	
	\item The fuel mass flow injected into the swirler is not affected by the pressure oscillations, as suggested by Arndt et al.~\cite{Arndt2015}.  Contrary to the global mass flux a convective delay of the fuel mass flux peak is already observable inside the swirler.
\end{enumerate}
\vspace{-5pt}
\blue
\blueF
In summary, the results clearly show that the LES considering heat transfer effects is able to predict the global flow field more accurately than the adiabatic simulation. Further, the highly unsteady flame dynamics are captured with sufficient precision by the non-adiabatic LES, which is indispensable for the prediction of combustion instabilities.
%employing an  enthalpy-dependent tabulated chemistry approach is able to predict the global flow field and contrary to the adiabatic simulation, also the highly unsteady flame dynamics with sufficient precision, which is indispensable for the prediction of combustion instabilities. 
\color{black}

%the results indicate that the LES with an enthalpy-dependent tabulated chemistry approach can accurately reproduce not only the global flow field characteristics but also the dynamics of the flame inside the combustor since the phase-conditioned flow field variables and reactive scalars compare fairly well with the experiments.
%, while adiabatic tables are not sufficient to capture the correct mixture field and flame position. 

%% file: Manuscript.bbl
\begin{thebibliography}{53}
\providecommand{\natexlab}[1]{#1}
\providecommand{\url}[1]{\texttt{#1}}
\expandafter\ifx\csname urlstyle\endcsname\relax
  \providecommand{\doi}[1]{doi: #1}\else
  \providecommand{\doi}{doi: \begingroup \urlstyle{rm}\Url}\fi

\bibitem[Huang and Yang(2009)]{Huang2009}
Y.~Huang and V.~Yang.
\newblock Dynamics and stability of lean-premixed swirl-stabilized combustion.
\newblock \emph{Prog. Energy Combust. Sci.}, 35\penalty0 (4):\penalty0
  293--364, 2009.
\newblock \doi{10.1016/j.pecs.2009.01.002}.

\bibitem[Gicquel et~al.(2012)Gicquel, Staffelbach, and Poinsot]{Gicquel2012}
L.~Y.~M. Gicquel, G.~Staffelbach, and T.~Poinsot.
\newblock Large eddy simulations of gaseous flames in gas turbine combustion
  chambers.
\newblock \emph{Prog. Energy Combust. Sci.}, 38\penalty0 (6):\penalty0
  782--817, 2012.
\newblock \doi{10.1016/j.pecs.2012.04.004}.

\bibitem[Oberleithner et~al.(2015)Oberleithner, Stöhr, Im, Arndt, and
  Steinberg]{Oberleithner2015}
K.~Oberleithner, M.~Stöhr, S.~H. Im, C.~M. Arndt, and A.~M. Steinberg.
\newblock Formation and flame-induced suppression of the precessing vortex core
  in a swirl combustor: Experiments and linear stability analysis.
\newblock \emph{Combust. Flame}, 162\penalty0 (8):\penalty0 3100--3114, 2015.
\newblock \doi{10.1016/j.combustflame.2015.02.015}.

\bibitem[Poinsot(2017)]{Poinsot2017}
T.~Poinsot.
\newblock Prediction and control of combustion instabilities in real engines.
\newblock \emph{Proc. Combust. Inst.}, 36\penalty0 (1):\penalty0 1--28, 2017.
\newblock \doi{10.1016/j.proci.2016.05.007}.

\bibitem[Schuller et~al.(2003)Schuller, Durox, and Candel]{Schuller2003}
T.~Schuller, D.~Durox, and S.~Candel.
\newblock {Self-induced combustion oscillations of laminar premixed flames
  stabilized on annular burners}.
\newblock \emph{Combust. Flame}, 135\penalty0 (4):\penalty0 525--537, 2003.
\newblock ISSN 00102180.
\newblock \doi{10.1016/j.combustflame.2003.08.007}.

\bibitem[Lieuwen and Yang(2006)]{Lieuwen2006}
T.~C. Lieuwen and V.~Yang.
\newblock \emph{{Combustion Instabilities In Gas Turbine Engines}}.
\newblock AAIA, Reston ,VA, 2006.
\newblock ISBN 978-1-56347-669-3.
\newblock \doi{10.2514/4.866807}.
\newblock URL \url{http://arc.aiaa.org/doi/book/10.2514/4.866807}.

\bibitem[Schadow and Gutmark(1992)]{Schadow1992}
K.~C. Schadow and E.~Gutmark.
\newblock {Combustion instability related to vortex shedding in dump combustors
  and their passive control}.
\newblock \emph{Prog. Energy Combust. Sci.}, 18\penalty0 (2):\penalty0
  117--132, 1992.
\newblock ISSN 03601285.
\newblock \doi{10.1016/0360-1285(92)90020-2}.

\bibitem[Palies et~al.(2011)Palies, Schuller, Durox, Gicquel, and
  Candel]{Palies2011}
P.~Palies, T.~Schuller, D.~Durox, L.~Y. Gicquel, and S.~Candel.
\newblock {Acoustically perturbed turbulent premixed swirling flames}.
\newblock \emph{Phys. Fluids}, 23\penalty0 (3), 2011.
\newblock ISSN 10706631.
\newblock \doi{10.1063/1.3553276}.

\bibitem[Boxx et~al.(2012)Boxx, Arndt, Carter, and Meier]{Boxx2012}
I.~Boxx, C.~M. Arndt, C.~D. Carter, and W.~Meier.
\newblock High-speed laser diagnostics for the study of flame dynamics in a
  lean premixed gas turbine model combustor.
\newblock \emph{Exp. Fluids}, 52:\penalty0 555--567, 2012.
\newblock ISSN 0723-4864.
\newblock \doi{10.1007/s00348-010-1022-x}.

\bibitem[Stöhr et~al.(2018)Stöhr, Oberleithner, Sieber, Yin, and
  Meier]{Stoehr2018}
M.~Stöhr, K.~Oberleithner, M.~Sieber, Z.~Yin, and W.~Meier.
\newblock Experimental study of transient mechanisms of bistable flame shape
  transitions in a swirl combustor.
\newblock \emph{J. Eng. Gas Turbines Power}, 140, 2018.
\newblock ISSN 0742-4795.
\newblock \doi{10.1115/1.4037724}.

\bibitem[Lieuwen and Zinn(1998)]{Lieuwen1998}
T.~Lieuwen and B.~T. Zinn.
\newblock {The role of equivalence ratio oscillations in driving combustion
  instabilities in low NOx gas turbines}.
\newblock \emph{Symp. (Int.) Combust.}, 27\penalty0 (2):\penalty0 1809--1816,
  1998.
\newblock ISSN 00820784.
\newblock \doi{10.1016/S0082-0784(98)80022-2}.

\bibitem[Meier et~al.(2007)Meier, Weigand, Duan, and
  Giezendanner-Thoben]{Meier2007}
W.~Meier, P.~Weigand, X.~R. Duan, and R.~Giezendanner-Thoben.
\newblock Detailed characterization of the dynamics of thermoacoustic
  pulsations in a lean premixed swirl flame.
\newblock \emph{Combust. Flame}, 150:\penalty0 2--26, 2007.
\newblock ISSN 0010-2180.
\newblock \doi{10.1016/j.combustflame.2007.04.002}.

\bibitem[Kim et~al.(2010)Kim, Lee, Quay, and Santavicca]{Kim2010}
K.~Kim, J.~Lee, B.~Quay, and D.~Santavicca.
\newblock Response of partially premixed flames to acoustic velocity and
  equivalence ratio perturbations.
\newblock \emph{Combust. Flame}, 157\penalty0 (9):\penalty0 1731--1744, 2010.
\newblock \doi{10.1016/j.combustflame.2010.04.006}.

\bibitem[{\'{C}}osi{\'{c}} et~al.(2015){\'{C}}osi{\'{c}}, Terhaar, Moeck, and
  Paschereit]{Cosic2015}
B.~{\'{C}}osi{\'{c}}, S.~Terhaar, J.~P. Moeck, and C.~O. Paschereit.
\newblock Response of a swirl-stabilized flame to simultaneous perturbations in
  equivalence ratio and velocity at high oscillation amplitudes.
\newblock \emph{Combust. Flame}, 162\penalty0 (4):\penalty0 1046--1062, 2015.
\newblock \doi{10.1016/j.combustflame.2014.09.025}.

\bibitem[Weigand et~al.(2006{\natexlab{a}})Weigand, Meier, Duan, Stricker, and
  Aigner]{Weigand2006a}
P.~Weigand, W.~Meier, X.~R. Duan, W.~Stricker, and M.~Aigner.
\newblock Investigations of swirl flames in a gas turbine model combustor.
\newblock \emph{Combust. Flame}, 144\penalty0 (1-2):\penalty0 205--224,
  2006{\natexlab{a}}.
\newblock \doi{10.1016/j.combustflame.2005.07.010}.

\bibitem[Meier et~al.(2006)Meier, Duan, and Weigand]{Meier2006}
W.~Meier, X.~R. Duan, and P.~Weigand.
\newblock Investigations of swirl flames in a gas turbine model combustor.
\newblock \emph{Combust. Flame}, 144\penalty0 (1-2):\penalty0 225--236, 2006.
\newblock \doi{10.1016/j.combustflame.2005.07.009}.

\bibitem[Weigand et~al.(2006{\natexlab{b}})Weigand, Meier, Duan, and
  Aigner]{Weigand2006}
P.~Weigand, W.~Meier, X.~Duan, and M.~Aigner.
\newblock Laser based investigations of thermo-acoustic instabilities in a lean
  premixed gas turbine model combustor.
\newblock In \emph{Volume 1: Combustion and Fuels, Education}, pages 237--245.
  {ASME}, 2006{\natexlab{b}}.
\newblock \doi{10.1115/gt2006-90300}.

\bibitem[Raman and Hassanaly(2019)]{Raman2019}
V.~Raman and M.~Hassanaly.
\newblock {Emerging trends in numerical simulations of combustion systems}.
\newblock \emph{Proc. Combust. Inst.}, 37\penalty0 (2):\penalty0 2073--2089,
  2019.
\newblock ISSN 15407489.
\newblock \doi{10.1016/j.proci.2018.07.121}.

\bibitem[Roux et~al.(2005)Roux, Lartigue, Poinsot, Meier, and
  B{\'{e}}rat]{Roux2005}
S.~Roux, G.~Lartigue, T.~Poinsot, U.~Meier, and C.~B{\'{e}}rat.
\newblock Studies of mean and unsteady flow in a swirled combustor using
  experiments, acoustic analysis, and large eddy simulations.
\newblock \emph{Combust. Flame}, 141\penalty0 (1-2):\penalty0 40--54, 2005.
\newblock \doi{10.1016/j.combustflame.2004.12.007}.

\bibitem[Franzelli et~al.(2012)Franzelli, Riber, Gicquel, and
  Poinsot]{Franzelli2012}
B.~Franzelli, E.~Riber, L.~Y.~M. Gicquel, and T.~Poinsot.
\newblock Large eddy simulation of combustion instabilities in a lean partially
  premixed swirled flame.
\newblock \emph{Combust. Flame}, 159\penalty0 (2):\penalty0 621--637, 2012.
\newblock \doi{10.1016/j.combustflame.2011.08.004}.

\bibitem[Galpin et~al.(2008)Galpin, Naudin, Vervisch, Angelberger, Colin, and
  Domingo]{Galpin2008}
J.~Galpin, A.~Naudin, L.~Vervisch, C.~Angelberger, O.~Colin, and P.~Domingo.
\newblock Large-eddy simulation of a fuel-lean premixed turbulent swirl-burner.
\newblock \emph{Combust. Flame}, 155\penalty0 (1-2):\penalty0 247--266, 2008.
\newblock \doi{10.1016/j.combustflame.2008.04.004}.

\bibitem[Wang et~al.(2014)Wang, Platova, Fröhlich, and Maas]{Wang2014}
P.~Wang, N.~Platova, J.~Fröhlich, and U.~Maas.
\newblock Large eddy simulation of the {PRECCINSTA} burner.
\newblock \emph{Int. J. Heat and Mass Transfer}, 70:\penalty0 486--495, 2014.
\newblock \doi{10.1016/j.ijheatmasstransfer.2013.11.025}.

\bibitem[See and Ihme(2015)]{See2015}
Y.~C. See and M.~Ihme.
\newblock Large eddy simulation of a partially-premixed gas turbine model
  combustor.
\newblock \emph{Proc. Combust. Inst.}, 35\penalty0 (2):\penalty0 1225--1234,
  2015.
\newblock \doi{10.1016/j.proci.2014.08.006}.

\bibitem[Gövert et~al.(2017)Gövert, Mira, Kok, V{\'{a}}zquez, and
  Houzeaux]{Goevert2017}
S.~Gövert, D.~Mira, J.~B.~W. Kok, M.~V{\'{a}}zquez, and G.~Houzeaux.
\newblock The effect of partial premixing and heat loss on the reacting flow
  field prediction of a swirl stabilized gas turbine model combustor.
\newblock \emph{Flow Turbul. Combust.}, 100\penalty0 (2):\penalty0 503--534,
  2017.
\newblock \doi{10.1007/s10494-017-9848-4}.

\bibitem[Fredrich et~al.()Fredrich, Jones, and Marquis]{Fredrich2019}
D.~Fredrich, W.~Jones, and A.~J. Marquis.
\newblock The stochastic fields method applied to a partially premixed swirl
  flame with wall heat transfer.
\newblock 205:\penalty0 446--456.
\newblock \doi{10.1016/j.combustflame.2019.04.012}.

\bibitem[Donini et~al.(2016)Donini, Bastiaans, van Oijen, and
  de~Goey]{Donini2016}
A.~Donini, R.~J.~M. Bastiaans, J.~A. van Oijen, and L.~P.~H. de~Goey.
\newblock A 5-d implementation of {FGM} for the large eddy simulation of a
  stratified swirled flame with heat loss in a gas turbine combustor.
\newblock \emph{Flow Turbul. Combust.}, 98\penalty0 (3):\penalty0 887--922,
  2016.
\newblock \doi{10.1007/s10494-016-9777-7}.

\bibitem[Benard et~al.(2019)Benard, Lartigue, Moureau, and Mercier]{Benard2019}
P.~Benard, G.~Lartigue, V.~Moureau, and R.~Mercier.
\newblock Large-eddy simulation of the lean-premixed {PRECCINSTA} burner with
  wall heat loss.
\newblock \emph{Proc. Combust. Inst.}, 37\penalty0 (4):\penalty0 5233--5243,
  2019.
\newblock \doi{10.1016/j.proci.2018.07.026}.

\bibitem[Kraus et~al.(2017)Kraus, Selle, Poinsot, Arndt, and
  Bockhorn]{Kraus2017}
C.~Kraus, L.~Selle, T.~Poinsot, C.~M. Arndt, and H.~Bockhorn.
\newblock Influence of heat transfer and material temperature on combustion
  instabilities in a swirl burner.
\newblock \emph{J. Eng. Gas Turbines Power}, 139, 2017.
\newblock ISSN 0742-4795.
\newblock \doi{10.1115/1.4035143}.

\bibitem[Kraus et~al.(2018)Kraus, Selle, and Poinsot]{Kraus2018}
C.~Kraus, L.~Selle, and T.~Poinsot.
\newblock Coupling heat transfer and large eddy simulation for combustion
  instability prediction in a swirl burner.
\newblock \emph{Combust. Flame}, 191:\penalty0 239--251, 2018.
\newblock ISSN 0010-2180.
\newblock \doi{10.1016/j.combustflame.2018.01.007}.

\bibitem[Tang and Raman(2021)]{Tang2021}
Y.~Tang and V.~Raman.
\newblock {Large eddy simulation of premixed turbulent combustion using a
  non-adiabatic, strain-sensitive flamelet approach}.
\newblock \emph{Combust. Flame}, 234:\penalty0 111655, 2021.
\newblock ISSN 00102180.
\newblock \doi{10.1016/j.combustflame.2021.111655}.
\newblock URL
  \url{https://www.sciencedirect.com/science/article/pii/S0010218021003989
  https://linkinghub.elsevier.com/retrieve/pii/S0010218021003989}.

\bibitem[Agostinelli et~al.(2021)Agostinelli, Laera, Boxx, Gicquel, and
  Poinsot]{Agostinelli2021}
P.~W. Agostinelli, D.~Laera, I.~Boxx, L.~Gicquel, and T.~Poinsot.
\newblock {Impact of wall heat transfer in Large Eddy Simulation of flame
  dynamics in a swirled combustion chamber}.
\newblock \emph{Combust. Flame}, 234:\penalty0 111728, 2021.
\newblock ISSN 15562921.
\newblock \doi{10.1016/j.combustflame.2021.111728}.
\newblock URL \url{https://doi.org/10.1016/j.combustflame.2021.111728}.

\bibitem[Tay-wo chong et~al.(2017)Tay-wo chong, Scarpato, and
  Polifke]{Tay-wo-chong2017}
L.~Tay-wo chong, A.~Scarpato, and W.~Polifke.
\newblock {LES combustion model with stretch and heat loss effects for
  prediction pf premixed flame characteristics and dynamics}.
\newblock pages 1--12. GT2017-63357 ASME, 2017.

\bibitem[Massey et~al.(2021)Massey, Chen, and Swaminathan]{Massey2021}
J.~C. Massey, Z.~X. Chen, and N.~Swaminathan.
\newblock {Modelling Heat Loss Effects in the Large Eddy Simulation of a Lean
  Swirl-Stabilised Flame}.
\newblock \emph{Flow, Turbul. Combust.}, 106\penalty0 (4):\penalty0 1355--1378,
  2021.
\newblock ISSN 15731987.
\newblock \doi{10.1007/s10494-020-00192-4}.
\newblock URL \url{https://doi.org/10.1007/s10494-020-00192-4}.

\bibitem[Tay-Wo-Chong and Polifke(2012)]{WoChong2012}
L.~Tay-Wo-Chong and W.~Polifke.
\newblock {LES}-based study of the influence of thermal boundary condition and
  combustor confinement on premix flame transfer functions.
\newblock In \emph{Proc. ASME Turbo Expo 2012}, pages 579--588. ASME, 2012.
\newblock \doi{10.1115/gt2012-68796}.

\bibitem[Arndt et~al.(2015)Arndt, Severin, Dem, Stöhr, Steinberg, and
  Meier]{Arndt2015}
C.~M. Arndt, M.~Severin, C.~Dem, M.~Stöhr, A.~M. Steinberg, and W.~Meier.
\newblock Experimental analysis of thermo-acoustic instabilities in a generic
  gas turbine combustor by phase-correlated {PIV}, chemiluminescence, and laser
  raman scattering measurements.
\newblock \emph{Exp. Fluids}, 56\penalty0 (4), 2015.
\newblock ISSN 0723-4864.
\newblock \doi{10.1007/s00348-015-1929-3}.

\bibitem[Meier et~al.(2016)Meier, Dem, and Arndt]{Meier2016}
W.~Meier, C.~Dem, and C.~M. Arndt.
\newblock Mixing and reaction progress in a confined swirl flame undergoing
  thermo-acoustic oscillations studied with laser raman scattering.
\newblock \emph{Exp. Therm. Fluid Sci.}, 73:\penalty0 71--78, 2016.
\newblock ISSN 0894-1777.
\newblock \doi{10.1016/j.expthermflusci.2015.09.011}.

\bibitem[Kraus et~al.(2016)Kraus, Harth, and Bockhorn]{Kraus2016}
C.~Kraus, S.~Harth, and H.~Bockhorn.
\newblock Experimental investigation of combustion instabilities in lean
  swirl-stabilized partially-premixed flames in single- and multiple-burner
  setup.
\newblock \emph{Int. J. Spray Combust. Dyn.}, 8:\penalty0 4--26, 2016.
\newblock ISSN 1756-8277.
\newblock \doi{10.1177/1756827715627064}.

\bibitem[Arndt et~al.(2017)Arndt, Stöhr, Severin, Dem, and Meier]{Arndt2017}
C.~M. Arndt, M.~Stöhr, M.~J. Severin, C.~Dem, and W.~Meier.
\newblock Influence of air staging on the dynamics of a precessing vortex core
  in a dual swirl gas turbine model combustor.
\newblock In \emph{53rd {AIAA}/{SAE}/{ASEE} Joint Propulsion Conference}. AIAA,
  2017.
\newblock \doi{10.2514/6.2017-4683}.

\bibitem[Arndt et~al.(2020)Arndt, Nau, and Meier]{Arndt2020}
C.~M. Arndt, P.~Nau, and W.~Meier.
\newblock Characterization of wall temperature distributions in a gas turbine
  model combustor measured by 2d phosphor thermometry.
\newblock \emph{Proc. Combust. Inst.}, pages 1867--1875, 2020.
\newblock \doi{10.1016/j.proci.2020.06.088}.

\bibitem[Lourier et~al.(2017)Lourier, Stöhr, Noll, Werner, and
  Fiolitakis]{Lourier2017}
J.-M. Lourier, M.~Stöhr, B.~Noll, S.~Werner, and A.~Fiolitakis.
\newblock Scale adaptive simulation of a thermoacoustic instability in a
  partially premixed lean swirl combustor.
\newblock \emph{Combust. Flame}, 183:\penalty0 343--357, 2017.
\newblock \doi{10.1016/j.combustflame.2017.02.024}.

\bibitem[OpenFOAM(2013)]{OpenFOAM}
OpenFOAM.
\newblock The open source cfd toolbox, openfoam, 2013.
\newblock URL \url{http://www.openfoam.com}.

\bibitem[van Oijen and de~Goey(2000)]{Vanoijen2000}
J.~A. van Oijen and L.~P.~H. de~Goey.
\newblock {Modelling of Premixed Laminar Flames using Flamelet-Generated
  Manifolds Modelling of Premixed Laminar Flames using Flamelet-Generated
  Manifolds}.
\newblock \emph{Combust. Sci. Technol.}, 161:\penalty0 113--137, 2000.
\newblock \doi{10.1080/00102200008935814}.

\bibitem[Popp et~al.(2015)Popp, Hunger, Hartl, Messig, Coriton, Frank, Fuest,
  and Hasse]{Popp2015}
S.~Popp, F.~Hunger, S.~Hartl, D.~Messig, B.~Coriton, J.~H. Frank, F.~Fuest, and
  C.~Hasse.
\newblock {LES} flamelet-progress variable modeling and measurements of a
  turbulent partially-premixed dimethyl ether jet flame.
\newblock \emph{Combust. Flame}, 162\penalty0 (8):\penalty0 3016--3029, 2015.
\newblock \doi{10.1016/j.combustflame.2015.05.004}.

\bibitem[Gierth et~al.(2018)Gierth, Hunger, Popp, Wu, Ihme, and
  Hasse]{Gierth2018}
S.~Gierth, F.~Hunger, S.~Popp, H.~Wu, M.~Ihme, and C.~Hasse.
\newblock Assessment of differential diffusion effects in flamelet modeling of
  oxy-fuel flames.
\newblock \emph{Combust. Flame}, 197:\penalty0 134--144, 2018.
\newblock \doi{10.1016/j.combustflame.2018.07.023}.

\bibitem[Popp et~al.(2021)Popp, Hartl, Butz, Geyer, Dreizler, Vervisch, and
  Hasse]{Popp2020}
S.~Popp, S.~Hartl, D.~Butz, D.~Geyer, A.~Dreizler, L.~Vervisch, and C.~Hasse.
\newblock {Assessing multi-regime combustion in a novel burner configuration
  with large eddy simulations using tabulated chemistry}.
\newblock \emph{Proc. Combust. Inst.}, 38\penalty0 (2):\penalty0 2551--2558,
  2021.
\newblock ISSN 15407489.
\newblock \doi{10.1016/j.proci.2020.06.098}.
\newblock URL \url{https://doi.org/10.1016/j.proci.2020.06.098
  https://linkinghub.elsevier.com/retrieve/pii/S1540748920301589}.

\bibitem[Ketelheun et~al.(2013)Ketelheun, Kuenne, and Janicka]{Ketelheun2013}
A.~Ketelheun, G.~Kuenne, and J.~Janicka.
\newblock Heat transfer modeling in the context of large eddy simulation of
  premixed combustion with tabulated chemistry.
\newblock \emph{Flow Turbul. Combust.}, 91\penalty0 (4):\penalty0 867--893,
  2013.
\newblock \doi{10.1007/s10494-013-9492-6}.

\bibitem[Steinhausen et~al.(2020)Steinhausen, Luo, Popp, Strassacker, Zirwes,
  Kosaka, Zentgraf, Maas, Sadiki, Dreizler, and Hasse]{Steinhausen2020}
M.~Steinhausen, Y.~Luo, S.~Popp, C.~Strassacker, T.~Zirwes, H.~Kosaka,
  F.~Zentgraf, U.~Maas, A.~Sadiki, A.~Dreizler, and C.~Hasse.
\newblock Numerical investigation of local heat-release rates and
  thermo-chemical states in side-wall quenching of laminar methane and dimethyl
  ether flames.
\newblock \emph{Flow Turbul. Combust.}, page 681–700, 2020.
\newblock \doi{10.1007/s10494-020-00146-w}.

\bibitem[Zschutschke et~al.(2017)Zschutschke, Messig, Scholtissek, and
  Hasse]{ulf}
A.~Zschutschke, D.~Messig, A.~Scholtissek, and C.~Hasse.
\newblock Universal laminar flame solver (ulf).
\newblock 2017.
\newblock \doi{10.6084/m9.figshare.5119855.v2}.
\newblock URL
  \url{https://figshare.com/articles/poster/ULF\_code\_pdf/5119855}.

\bibitem[Smith et~al.()Smith, Golden, Frenklach, Moriarty, Eiteneer,
  Goldenberg, Bowman, Hanson, Song, Gardiner, Lissianski, and Qin]{grimech3}
G.~P. Smith, D.~M. Golden, M.~Frenklach, N.~W. Moriarty, B.~Eiteneer,
  M.~Goldenberg, C.~T. Bowman, R.~K. Hanson, S.~Song, W.~C. Gardiner, J.~V.~V.
  Lissianski, and Z.~Qin.
\newblock Gri-mech 3.0.
\newblock URL
  \url{http://combustion.berkeley.edu/gri-mech/version30/text30.html}.

\bibitem[Nicoud et~al.(2011)Nicoud, Toda, Cabrit, Bose, and Lee]{Nicoud2011}
F.~Nicoud, H.~B. Toda, O.~Cabrit, S.~Bose, and J.~Lee.
\newblock Using singular values to build a subgrid-scale model for large eddy
  simulations.
\newblock \emph{Phys. Fluids}, 23, 2011.
\newblock \doi{10.1063/1.3623274}.

\bibitem[Colin et~al.(2000)Colin, Ducros, Veynante, and Poinsot]{Colin2000}
O.~Colin, F.~Ducros, D.~Veynante, and T.~Poinsot.
\newblock {A thickened flame model for large eddy simulations of turbulent
  premixed combustion}.
\newblock \emph{Phys. Fluids}, 12\penalty0 (7):\penalty0 1843--1863, 2000.
\newblock ISSN 1070-6631.
\newblock \doi{10.1063/1.870436}.
\newblock URL \url{http://aip.scitation.org/doi/10.1063/1.870436}.

\bibitem[Charlette et~al.(2002)Charlette, Meneveau, and
  Veynante]{Charlette2002}
F.~Charlette, C.~Meneveau, and D.~Veynante.
\newblock A power-law flame wrinkling model for les of premixed turbulent
  combustion part i: non-dynamic formulation and initial tests.
\newblock \emph{Combust. Flame}, 131:\penalty0 159--180, 2002.
\newblock ISSN 0010-2180.
\newblock \doi{10.1016/s0010-2180(02)00400-5}.

\bibitem[Popp et~al.(2019)Popp, Kuenne, Janicka, and Hasse]{Popp2019}
S.~Popp, G.~Kuenne, J.~Janicka, and C.~Hasse.
\newblock An extended artificial thickening approach for strained premixed
  flames.
\newblock \emph{Combust. Flame}, 206:\penalty0 252--265, 2019.
\newblock ISSN 0010-2180.
\newblock \doi{10.1016/j.combustflame.2019.04.047}.

\end{thebibliography}
